

\input phyzzx

%
\catcode`\@=11 
\def\papers{\papersize\headline=\paperheadline\footline=\paperfootline}
\def\papersize{\hsize=40pc \vsize=53pc \hoffset=0pc \voffset=1pc
   \advance\hoffset by\HOFFSET \advance\voffset by\VOFFSET
   \pagebottomfiller=0pc
   \skip\footins=\bigskipamount \normalspace }
\catcode`\@=12 
\papers

\def\to{\rightarrow}

\vsize=21.5cm
\hsize=15.cm

\tolerance=500000
\overfullrule=0pt

\Pubnum={PUPT-1482 \cr
hep-th@xxx/9407151 \cr
July 1994}

\date={}
\pubtype={}
\titlepage
\title{GRAVITATIONALLY DRESSED CONFORMAL FIELD THEORY\break
AND EMERGENCE OF LOGARITHMIC OPERATORS}
\author{{
Adel~Bilal}\foot{
 on leave of absence from (also address after August 15, 1994)
Laboratoire de Physique Th\'eorique de l'Ecole
Normale Sup\'erieure, 24 rue Lhomond, 75231
Paris Cedex 05, France
(unit\'e propre du CNRS)\nextline
e-mail: bilal@puhep1.princeton.edu
}}
\andauthor{{Ian I. Kogan}\foot{
 On  leave of absence
from ITEP,
 B.Cheremyshkinskaya 25,  Moscow, 117259,     Russia. \nextline
 Address after August 15, 1994: Department of Theoretical
 Physics, 1 Keble Road, Oxford\nextline
OX1 3NP, UK\nextline
e-mail:kogan@puhep1.princeton.edu}}
\address{\it Joseph Henry Laboratories\break
Princeton University\break
Princeton, NJ 08544, USA}

\vskip 3.mm
\abstract{
We study correlation functions in two-dimensional conformal
field theory coupled to induced gravity in the light-cone
gauge. Focussing on the fermion four-point function, we
display an unexpected non-perturbative singularity structure:
coupling to gravity {\it qualitatively} changes the perturbative
$(x_1-x_2)^{-1}(x_3-x_4)^{-1}$ singularity into a logarithmic one
plus a non-singular piece. We argue that this is related to the
appearence of new logarithmic operators in the gravitationally
dressed operator product expansions.

We also show some evidence that non-conformal but integrable
models may remain integrable when coupled to gravity.
}

\endpage
\pagenumber=1

 \def\PL #1 #2 #3 {Phys.~Lett.~{\bf #1} (#2) #3}
 \def\NP #1 #2 #3 {Nucl.~Phys.~{\bf B  #1} (#2) #3}
 \def\PR #1 #2 #3 {Phys.~Rev.~{\bf #1} (#2) #3}
 \def\PRL #1 #2 #3 {Phys.~Rev.~Lett.~{\bf #1} (#2) #3}
 \def\CMP #1 #2 #3 {Comm.~Math.~Phys.~{\bf #1} (#2) #3}
 \def\IJMP #1 #2 #3 {Int.~J.~Mod.~Phys.~{\bf #1} (#2) #3}
 \def\JETP #1 #2 #3 {Sov.~Phys.~JETP.~{\bf #1} (#2) #3}
 \def\PRS #1 #2 #3 {Proc.~Roy.~Soc.~{\bf #1} (#2) #3}
 \def\IM #1 #2 #3 {Inv.~Math.~{\bf #1} (#2) #3}
 \def\JFA #1 #2 #3 {J.~Funkt.~Anal.~{\bf #1} (#2) #3}
 \def\LMP #1 #2 #3 {Lett.~Math.~Phys.~{\bf #1} (#2) #3}
 \def\IJMP #1 #2 #3 {Int.~J.~Mod.~Phys.~{\bf #1} (#2) #3}
 \def\FAA #1 #2 #3 {Funct.~Anal.~Appl.~{\bf #1} (#2) #3}
 \def\AP #1 #2 #3 {Ann.~Phys.~{\bf #1} (#2) #3}
 \def\MPL #1 #2 #3 {Mod.~Phys.~Lett.~{\bf A #1} (#2) #3}

\def\d{\partial}
\def\dt{\partial_t}
\def\f{\phi}
\def\ix{\int {\rm d}^2 x \, }
\def\e{\epsilon}

\def\a{\alpha}
\def\g{\gamma}
\def\G{\Gamma}
\def\xm{x^-}
\def\ym{y^-}
\def\zm{z^-}
\def\wm{w^-}
\def\xjm{x^-_j}
\def\xp{x^+}
\def\yp{y^+}
\def\zp{z^+}
\def\wp{w^+}
\def\xjp{x^+_j}
\def\tb{{\bar t}}
\def\D{\Delta}
\def\rmd{{\rm d}}

\def\la{\langle}
\def\ra{\rangle}

\def\P{\Psi}
\def\p{\psi}
\def\dd #1 #2{{\delta #1\over \delta #2}}

\def\lg{\la\la}
\def\rg{\ra\ra}

\REF\POL{A. Polyakov, \MPL 2 1987 893 \nextline
 V. Knizhnik, A. Polyakov and A. Zamolodchikov, \MPL 3 1988 819}

In this letter we consider the gravitational dressing in a
 two-dimensional  conformal field theory  coupled to  two-dimensional
 quantum gravity. This problem is important  not only for  conformal
 models on random surfaces and non-critical strings but also
  as a starting point for understanding  the effect of gravitational
 dressing in  renormalizable but not conformally invariant models.
 One of the interesting possibilities is the appearence of new
 states in the theory due to the inclusion of gravity. This phenomenon
 is   non-perturbative
and it is the purpose of this letter to explore it to some extent and
to  show that the new states  indeed appear. To find these states
 and examine
 their unusual properties we shall study the gravitationally dressed
 $4$-point function using the light-cone formulation of two-dimensional
 quantum gravity [\POL].

\REF\HOU{A. Polyakov, in  {\it Les Houches 1988: Fields, Strings and
Critical Phenomena}, Elsevier Science Publishers 1989}
\REF\KKP{I. Klebanov, I. Kogan and A. Polyakov, \PRL 71 1993 3243}
To begin with, following [\HOU-\KKP], we derive a differential equation for
the gravitationally dressed fermion $n$-point function. In light-cone
gauge, gravity is taken into account by adding $\ix h_{++}(x)
T_{--}(x)$ to the action so that the gravitationally dressed
correlation functions are
$$\lg \f_1(x_1)\ldots \f_n(x_n)\rg
=\int {\cal D}h_{++} \la \f_1(x_1)\ldots \f_n(x_n)\ra \exp\left(
i\ix h_{++}(x) T_{--}(x)\right)\ .
\eqn\i$$
If $\f_j$ has conformal dimension $\D_j$, i.e. if under a
reparametrization
$$\delta_\e \f_j=\e_+\d_-\f_j+\D_j (\d_-\e_+)\f_j
\eqn\ii$$
then the corresponding integrated Ward identity is [\HOU]
$$\eqalign{
&\g \lg h_{++}(z) \f_1(x_1)\ldots \f_n(x_n)\rg +
\sum_{j=1}^n \left[ {(\zm-\xjm)^2\over \zp-\xjp}\, {\d\over \d\xjm}
-2\D_j\, {\zm-\xjm\over \zp-\xjp}\right]
\lg \f_1(x_1)\ldots \f_n(x_n)\rg \cr
&=0\ .}
\eqn\iii$$
Here $\D_j$ are the conformal dimensions in the presence of gravity as
given by the gravitationally dressed two-point functions, and $\g$ is
related to the level $k$ of the gravitational $Sl(2,{\bf R})$ current
algebra [\POL] by
$$\g=k+2={1\over 12}\left( c-13-\sqrt{(c-1)(c-25)}\right)
\eqn\iv$$
where $c$ is the total central charge of the matter coupling to
gravity.

Using now the quantum equations of motion (i.e. Schwinger-Dyson
equations) for a left-handed fermion in the presence of gravity
$$\d_+\p=:h_{++}\d_-\p:+\D:(\d_-h_{++})\p:
\eqn\iva$$
one obtains for the $n$ fermion correlation function
$$\eqalign{
\left\{ \g{\d\over \d \zp}+\sum_{j=2}^n \left[ {(\zm-\xjm)^2\over
\zp-\xjp} {\d\over \d\zm}{\d\over \d\xjm} -2\D {\zm-\xjm\over \zp-\xjp}
\left({\d\over \d\zm}-{\d\over \d\xjm}\right)
-{2\D^2\over \zp-\xjp}
\right]\right\}&\cr
\lg \p(z) \p(x_2)\ldots \p(x_n)\rg= 0& \ .\cr}
\eqn\v$$
One may check that this gives the two-point function correctly as
$$\lg \p(z)\p(x)\rg={1\over (\zm-\xm)^{2\D}\, (\zp-\xp)^{2\D-1} }
\eqn\vi$$
with
$$\D-{1\over 2}=-{\D(1-\D)\over k+2}\ .
\eqn\vii$$

We now concentrate on the fermion four-point function. Conformal
invariance implies
$$\eqalign{
G_4(w,x,y,z)&\equiv \lg\p_i(w)\p_i(x)\p_j(y)\p_j(z)\rg\cr
&={f(t^-,t^+)\over (\wm-\xm)^{2\D}\, (\wp-\xp)^{2\D-1}\,
(\ym-\zm)^{2\D}\, (\yp-\zp)^{2\D-1} }\ .\cr}
\eqn\viii$$
We have added colour indices for the fermions, and if $i\ne j$, then
obviously $f=1+{\cal O}(1/\g)$. The anharmonic ratio $t$ is given as
usual by
$$t\equiv t^-={(\wm-\ym)(\xm-\zm)\over (\wm-\zm)(\xm-\ym)}
\eqn\ix$$
and similarly for $\tb\equiv t^+$. Inserting the ansatz \viii\ into
eq. \v\ with $n=4$ leads after some algebra to
$$\left[ \g\tb\d_{\tb} +{1-t\over 1-\tb}\, (\tb-t)\, \dt\, t\dt
+(1-4\D) t\dt +2\D^2\, {t+1\over t-1}\right] f(t,\tb)=0\ .
\eqn\x$$
Note that for $\D=1$ this reproduces the equation derived in [\KKP]
for the four-current correlation function.

This partial non-linear differential equation \x\ seems too
complicated to be solved in full generality. In principle, it can be
solved in a perturbative series order by order in $1/\g\sim 1/c$. To
fix the integration ambiguity it is useful to compare with a direct
Feynman diagram computation for the four-point function. The latter
gives (for colour indices $i\ne j$)
$$f=1-{1\over 2\g}\, {t+1\over t-1}\log t\tb +{\cal O}(1/\g^2)
=1-{2\D^2\over \g}\, {t+1\over t-1}\log t\tb +{\cal O}(1/\g^2) \ .
\eqn\xi$$
\REF\BK{A.Bilal and I. Kogan, work in progress}
Although not obvious on the form \xi\ the momentum space four-point
function (with the external legs removed)
vanishes at this order when the fermions are on shell.
Indeed, it is easy to see that it
 is proportional to
$$
{1\over \gamma}  (p + p')_{-}(q + q')_{-}{(p-p')_{+}\over (p-p')_{-}^{3}}
$$
 where $p,~p'$ and $q,~q'$ are the initial and final momenta
 of the first and second fermions.  For left fermions one has the
 on-shell condition $p_{+} = p'_{+}= 0$ (and the same for $q$)
   and one gets no contribution to the  $S$-matrix at this order. We
have verified that this remains true at the next order, including two
graviton exchanges. In other words, up to this order, the $S$-matrix for
gravitational scattering of two left fermions is unity. If this turns
out to be true at all orders one would have an interesting
situation: Consider the Gross-Neveu model coupled to gravity. In the
light-cone gauge only the left fermions couple to gravity. Left-left
scattering would then be elastic, as is right-right scattering
(anyhow) and left-right scattering (by kinematics). Thus any
two-particle scattering would be elastic. This is an encouraging
result to speculate that the integrability of the Gross-Neveu model
might survive coupling to gravity. One can even try to formulate
a gravitationally  dressed Bethe-Ansatz  which may help to solve
 exactly integrable models coupled to induced gravity. Details will
be given elsewhere [\BK].

The main difficulty in solving the differential equation \x\ is
 the factor $1/ (1-\tb)$. If one considers the vicinity of $t=1$,
this difficulty disappears, and \x\ becomes
$$\left[ \g\tb\d_{\tb} +(t-1)\dt\, t\dt
+(1-4\D) \dt + {4\D^2\over t-1}\right] f_1(t,\tb)=0
\eqn\xii$$
where the subscript $1$ on $f$ is to remind us that $f\sim f_1$
only in the vicinity of $t=1$. Equation \xii\ can be solved
exactly. First, perturbation in $1/\g$ leads to (matching with \xi)
$$f_1(t,\tb)=\sum_{n=0}^\infty {[(2\D)_n]^2\over n!}\, g^n\quad,
\quad g=-{\log \tb\over \g (t-1)}
\eqn\xiii$$
where $a_n\equiv a(a+1)(a+2)\ldots (a+n-1)$. At each order in $1/\g$
one can of course replace $\log \tb$ by $\log t\tb$ in $g$, as
suggested by \xi.\foot{
 Then $f_1$ is no longer an exact solution of
\xii\ but an exact solution of another equation, differing from \xii\
only by higher order terms in $(t-1)$, just as \xii\ differs from \x\ by
higer order terms in $(t-1)$ anyhow.}
The series \xiii\ has zero radius of convergence, but its Borel
transform can be recognized as the hypergeometric function
$$B[f_1](z)=\sum_{n=0}^\infty {[(2\D)_n]^2\over n! n!}\, z^n
=F(2\D, 2\D,1;z)\ .
\eqn\xiv$$
Inverting the Borel transform gives the resummed function $f_1(g)$ in
terms of the Whittaker function. Alternatively, one can directly
observe that \xiii\ coincides up to an overall factor with the
asymptotic expansion of the Whittaker function. Hence
\REF\GR{I.S. Gradshteyn and I.M. Ryzhik, {\it Table of integrals, series and
products}, Academic Press, 1980}
[\GR]
$$f_1(g)=\left(-1/ g\right)^{2\D}\P(2\D,1;-1/ g)
=\left(-1/ g\right)^{2\D-1/2}\, e^{-1/2g}\,
 W_{1/2-2\D,0}(-1/ g)\ .
\eqn\xv$$
Here $W$ is the Whittaker function and $\P$ is a solution to the
degenerate hypergeometric equation [\GR]. Indeed, although the
equation \xii\ has many solutions, if one uses an ansatz with $f_1$
only depending on $t$ and $\tb$ through $g$, then equation \xii\
becomes
$$\left\{ g^2{\rmd^2\over \rmd g^2} +[(1+4\D)g-1] {\rmd\over \rmd g}
+4\D^2\right\} f_1(g) =0\ .
\eqn\xvi$$
Setting $f_1(g)=(-1/g)^{2\D}\, u(-1/g)$ one sees that $u(x)$ satisfies
the degenerate hypergeometric equation
$$xu''(x)+(b-x)u'(x)-a u(x)=0
\eqn\xvii$$
with $b=1$ and $a=2\D$. Perturbation theory has told us which of the
two independent solutions to choose, namely $u(x)=\P(2\D,1;x)$.

Having the perfectly non-perturbative expression \xv\ for $f_1(g)$,
we can now investigate its behaviour for large $g$, which is just the
series expansion of $\P(2\D,1;x)$ for small $x$ [\GR]:
$$f_1(g)=\left(-{1\over g}\right)^{2\D}\sum_{k=0}^\infty
{\G(2\D+k)\over [k! \G(2\D)]^2}
\left[ 2\p(k+1)-\p(2\D+k)-\log\left(-{1\over g}\right) \right]
\left(-{1\over g}\right)^k
\eqn\xviii$$
where $\p(x)=\G'(x)/\G(x)$. Recall that $g=-\log\tb/[\g(t-1)]$, hence
large $g$ means $t\to 1$ (for fixed $\tb$), so this is the limit
where $f\sim f_1$. Although \xviii\ is the exact asymptotic for
$f_1$, it gives only the leading order for $f$ as $t\to 1$:
$$f(t,\tb)\, \sim \,
\left( {\g(t-1)\over \log\tb}\right)
{1\over \G(2\D)} \left[ \p(1)-\log\left( {\g(t-1)\over \log\tb}\right)
+{\cal O}\left( (t-1), (t-1)\log (t-1)\right) \right] \ .
\eqn\xix$$

What does this mean for the fermion four-point function \viii?
Since $t-1={(\wm-\xm)(\ym-\zm)\over (\wm-\zm)(\xm-\ym)}$, one has
$t\to 1$ if either $\wm\to\xm$ or $\ym\to\zm$, i.e. when two fermion
operators of the same colour approach each other. Inserting \xix\
into \viii\ then gives
$$\eqalign{G_4(w,x,y,z)\, \sim&\,
{\g^{2\D}\over \G(2\D)}\left[
(\wm-\zm)(\xm-\ym)\right]^{-2\D}\left[\log\tb\, \right]^{-2\D}
\left[(\wp-\xp)(\yp-\zp)\right]^{1-2\D}\cr
&\times \left[ \p(1)+\log\left({\log\tb\over \g}\right)
-\log {(\wm-\xm)(\ym-\zm)\over (\wm-\zm)(\xm-\ym)} \right] \ . \cr}
\eqn\xx$$
It is important to realize that we work in Minkowski space so that we
can take $t\to 1$, keeping $\tb\ne 1$ fixed. Rather surprisingly, the
four-point function \xx\ no longer contains the perturbative
singularity $\sim (\wm-\xm)^{-2\D} (\ym-\zm)^{-2\D}$, but resumming
the series has transformed it into a logarithmic singularity, plus a
non-singular part !

Mathematically, the origin of the logarithm can be traced to the
degenerate hypergeometric equation \xvii\ satisfied by $u(x)=x^{-2\D}
f_1(-1/x)$. For generic  parameter $b$ it has two independent
solutions [\GR] $\Phi(a,b;x)$ and $x^{1-b}\Phi(a+1-b,2-b;x)$.
Obviously, for $b\to 1$ the second solution generates $\log x\,
\Phi(a,b;x)$, among others. This is a well-known phenomenon in the
theory of ordinary linear differential equations.

Physically however, it was quite unexpected that turning on gravity
($1/\g \ne 0$), even infinitesimally weakly, completely changes the
singularity structure: this is a truely non-perturbative
phenomenon.

\REF\RS {L. Rozansky and H.Saleur, \NP 376 1992 441}
\REF\GUR{V.Gurarie, \NP 410 1993 535}
The appearence of logarithms in  correlation functions had been
 noticed before in [\RS] where the WZW model based on the supergroup
 GL$(1,1)$  was discussed. Later these logarithms were discussed in
the $c = -2$ model and other nonunitary non-minimal models  in [\GUR]. There
  it has been argued that the emergence of the logarithms in
 correlation functions are due to the existence of new operators in
 the operator product expansion with a new, ``logarithmic" behaviour.
The anomalous dimensions of these new operators are degenerate with those
of the usual primary operators.
 Then one no longer can completely diagonalize the Virasoro operator $L_{0}$
 and the new operators together with the  standard ones are the basis of
 the Jordan cell for $L_{0}$.
More precisely,
in the case of two
 operators $\tilde{O}_{n}$ and $O_{n}$ with degenerate anomalous dimensions
$\Delta_{n}$ the operator product expansion now takes the form
$$
\phi(x) \phi(0) = \sum_{n} x^{\Delta_{n} - 2\Delta_{\phi}}
\left[\tilde{O}_{n} + \ldots + \log (x) O_{n} + \ldots \right]\ .
\eqn\xxii$$
These operators also have unusual OPEs with the stress-energy tensor:
$$T(z) \tilde O_n(0) = {\D_n\over z^2}\tilde O_n(0)+{1\over z^2} O_n(0)
+{1\over z}\d_z\tilde O_n(0)
\eqn\xxiia $$
in particular
$$
L_{0} | O_{n}\ra = \Delta_{n} | O_{n}\ra,~~~~~~~   L_{0}|\tilde{O}_{n}\ra
 =
  \Delta_{n}|\tilde{O}_{n}\ra + |O_n\ra
\eqn\xxiii$$
This allows us to have logarithmic terms in correlation functions without
spoiling the conformal invariance.
Using these rules one can extract information about the OPEs in the
gravitaionally
 dressed theory from the correlation function \xx.
Obviously, as far as the left dimensions are concerned, eq. \xx\
corresponds to $\D_n-2\D_\psi=0$. More details will be given elsewhere.

\REF\POL{J. Polchinski, \NP 346 1990  253}
Let us note that  in the $c=1$ string one also encounters an
analogous phenomenon:
As shown in [\POL], in the linear approximation, the
 tachyon  field ${\cal T}$ obeys the
 following equation
$$
-\partial^{2}_{t}{\cal T}
 + \partial^{2}_{\phi}{\cal T} + 2\sqrt{2} \partial_{\phi}{\cal T} +
 2{\cal T } + .. = 0
\eqn\xxiv$$
where $\phi$ is the Liouville field.  For solutions independent of $t$
 the characteristic equation $\lambda^{2} +  2\sqrt{2} \lambda
 + 2 =  (\lambda + \sqrt{2})^{2} = 0$ has degenerate roots, and the
 two solutions are $\exp(-\sqrt{2}\phi)$ and $\phi\exp(-\sqrt{2}\phi)$.
 Considering correlation functions of the type
$<{\cal T}(z_1) {\cal T} (z_2) ...>$
 one gets structures  like  $z^\a$ and $\log z\  z^\a$,
  revealing again in the conformal gauge signs of
the logarithmic terms we found in the light-cone
 formulation of 2-dimensional gravity.
 In both cases one has degeneracy of the anomalous dimensions of some
 conformal fields.

In conclusion, we have derived the exact differential equation \x\ for
the gravitationally dressed fermion four-point function. In the vicinity
of $t\to 1$ this equation simplifies, and the simplified equation \xii\
could be solved to all orders in $1/\g\sim 1/c$. We could resum the divergent
perturbation series, and thus study the singularity structure of the
four-point function for $t\to 1$ (non-perturbative regime). Rather
surprisingly, non-perturbative effects of gravity have changed the
$(\wm-\xm)^{-1}(\ym-\zm)^{-1}$ singularity into a logarithmic one plus
a regular term, thus indicating the appearence of new logarithmic operators.

\ack

We are grateful to D. Gross, I. Klebanov and A. Polyakov for sharing their
insights. The work of I.I.K. was supported by NSF grant No. PHY90-21984.

\refout
\end